\documentstyle[preprint,aps]{revtex}

\makeatletter
\def\lsim{\mathrel{\mathpalette\subsim@align<}}
\def\gsim{\mathrel{\mathpalette\subsim@align>}}
\def\subsim@align#1#2{\lower.6ex\vbox{\baselineskip\z@skip\lineskip\z@
\ialign{$\m@th#1\hfil##\hfil$\crcr#2\crcr\sim\crcr}}}
\makeatother

\begin{document}
\draft
\preprint{9411}

\title{
Power moments and scaling properties of nuclear emulsion data
   }

\author{S.\ J.\ Lee
  \footnote{electronic address: ssjlee@nms.kyunghee.ac.kr \ \& \
                                sjlee@ruthep.rutgers.edu}
 }
\address{
 Department of Physics, College of Natural Sciences \\
 Kyung Hee University, Yongin, Kyungkido \ 449-701, Korea
     }
\author{and}
\author{A.\ Z.\ Mekjian
  \footnote{electronic address: mekjian@ruthep.rutgers.edu}
 }
\address{
 Deparment of Physics \& Astronomy\\
Rutgers University, Piscataway, New Jersey \ 08855-0849, U.S.A.
   }

\date{\today}

\maketitle

\begin{abstract}
Various features of the charge yield in heavy ion collisions are studied.
Considered here are the nuclear emulsion data of $^{84}$Kr, $^{131}$Xe,
$^{197}$Au, $^{238}$U in the energy range of 1 GeV/A.
Mean charge yields and multiplicity distribution of these data indicate
that the nuclear fragmentation at this energy is a multiprocess phenomena.
Power moments of the charge distribution exhibit a scale invariance.
While small nuclei (Kr and Xe) show a nuclear size dependence,
large nuclei (Au and U) exhibit scale independence in their charge
distribution.
\end{abstract}


\narrowtext

\section{Introduction}

The purpose of this paper is to explore scale invariant characteristics
appearing in nuclear fragmentation processes. The scale invariance
can be studied in terms of intermittency
\cite{e1,e2,e3,e5,e6,e7,e8,intm} and anomalous dimensions
\cite{intm,fractal,map,prlsb}.
These quantities for a given distribution are defined through
the bin size dependence of factorial moments or power moments.
Scale invariance indicates that the system has no inherent length scale.
Another way of saying that the system has no inherent scale is to say
all length scales are present \cite{otherx}.
The scale invariance may be related with a chaotic behavior of the
system. In quantum chaos, chaos is characterized by the level spacing
distribution. If the system is chaotic, then the level spacing has a Wigner
distribution instead of a Poisson distribution.

We have analyzed the charge distribution for Au emulsion data at 0.99
GeV/A \cite{intm,prlsb}.
Some indications of scale invariance behavior in the emulsion data is found.
The systematic analysis for Kr, Xe, Au, and U in the 1 GeV/A energy
region would be an interesting problem concernig the scale invariance.
The mean behavior and the power moments of the charge distribution
of these data will be analyzed and compared with a simple exactly
solvable nuclear fragmentation model \cite{otherx,general,mekjian}.

The simple fragmentation model will first be briefly reviewed and
the distribution will be reformulated in terms of a dimensionless variable.
Using this dimensionless variable, the mean behavior and the power moments
will be analyzed in our study of scale invariance and anomalous
dimensions.

\section{Review of fragmentation model and power moments}

A method for investigating the fragmentation of an object made of
$A$ elements into pieces of smaller size is developed
\cite{otherx,general,mekjian}.
Specifically, the initial $A$ elements end up in clusters or groups
of varying sizes characterized by the number of elements $i$ in
the cluster. The number of objects of size $i$ is called $n_i$
and a sum rule $A = \sum_{i=1}^A i n_i$ is a constraint.
Weighting each partition $\vec n = \{n_i\} = (n_1, n_2, ..., n_A)$ of $A$
by Cauchy's number $M_2(A,\vec n) = A!/[\prod_{j=1}^A j^{n_j} n_j!]$,
and weighting each cluster of size $i$ by a parameter $x_i$,
we have the corresponding canonical partition function:
\begin{eqnarray}
 Q_A(\vec x) = \sum_{\{n_i\}_A} M_2(A,\vec n) \prod_{i=1}^A [x_i^{n_i}]
       = \sum_{\{n_i\}_A} A! \prod_{i=1}^A
          \left[\frac{x_i^{n_i}}{i^{n_i}n_i!}\right] .  \label{qaxk}
\end{eqnarray}
Here $\vec x = (x_1, x_2, ..., x_A)$ and the sum is over all the possible
partitions $\vec n$ of a fixed $A$.
The mean number of cluster size $k$ for this case is given simply by
\begin{eqnarray}
 <n_k> = \left. \sum_{\{n_i\}_A} n_k M_2(A,\vec n) \prod_{i=1}^A [x_i^{n_i}]
            \right/ Q_A(\vec x)
       = \left[\frac{x_k}{k}\right] \frac{A!}{(A-k)!}
         \frac{Q_{A-k}(\vec x)}{Q_A(\vec x)} .   \label{nkxk}
\end{eqnarray}
The Cauchy's number $M_2(A,\vec n)$ contains the microstate counting
rule which is based on a combinatorial analysis for a partition
$\vec n$ of $A$ elements.
The counting factor $M_2(A,\vec n)$ also represents the number of
permutations of $A$ objects in a specific cycle class with $n_1$ unit
cycle, $n_2$ cycles of length 2, etc.
The canonical partition function $Q_A(\vec x)$
which will be used to describe a fragmentation process
is also the cycle indicator of the permutation or symmetric group.

Depending on the size $i$ dependence of the cluster weight
parameter $x_i$, the fragmentation model characterized by the
canonical partition function of Eq.(\ref{qaxk}) can describe
the mean distribution in various systems
\cite{otherx,general} by Eq.(\ref{nkxk}).
Specifically, a particular choice $x_i = x X_1^i$ reduces
Eqs.(\ref{qaxk}) and (\ref{nkxk}) into
\begin{eqnarray}
 Q_A(\vec x) &=& X_1^A Q_A(x) = X_1^A \frac{\Gamma(x+A)}{\Gamma(x)} ,
          \label{qax} \\
 <n_k> &=& \left[\frac{x}{k}\right] \frac{A!}{(A-k)!}
          \frac{\Gamma(x+A-k)}{\Gamma(x+A)} .   \label{nkx}
\end{eqnarray}
The $\Gamma$ is gamma function.
Notice here that $X_1$ disappears in $<n_k>$ because of the
constraint $\sum_i ~i n_i = A$.
This simple exact solution has been used in describing nuclear
fragment distributions produced in various proton-nucleus and
nucleus-nucleus collisions \cite{frgft}.
This simple formular of Eq.(\ref{nkx}) gives as good of a fit to the
data as complicated Monte-Carlo calculations \cite{pronuc}.
Depending on the value of $x$, the distribution has a behavior which is
a U shape curve (characteristics of a low energy or low
temperature system) or has an exponential fall off (high temperature)
with increasing cluster size \cite{general,frgft}.
This model also produces the scale invariance (self-similarity)
and critical exponents
appearing in the wide range of phenomena \cite{otherx}.
Thermodynamic argument relates the single parameter $x$
to the thermodynamic partition function
of a cluster at a given temperature and freeze out volume
\cite{general,frgft}.

Considering the mass distribution $p(k) = k n_k /A$,
we can relate our fragmentation model with a level spacing
distribution and with a ``devil's'' staircase for the cummulative
mass distribution \cite{intm}.
The level spacing for a specific partition $\vec n$ is defined
as $m_k = k n_k$ and the spacing distribution $P(m_k)$ as the probability
of having the spacing value $m_k = k n_k$ in the canonical ensemble.
The Wigner type level spacing distribution indicates quantum chaos
and the Poisson distribution indicates nonchaotic behavior.
The cummulative mass defined by
\begin{eqnarray}
 M(\kappa) = \int_0^\kappa dk ~k n_k ~=~ \int_0^\kappa dk ~m_k ,  \label{cumas}
\end{eqnarray}
form a ``devil's'' staircase for each partition.

We also have looked at various correlations and fluctuations
in the number of clusters $n_k$ or in the mass distribution
$p(k) = k n_k /A$ of cluster sizes in this simple model
\cite{intm,prlsb,general}.
The fluctuations and the correlations in this model are given simply by
\begin{eqnarray}
 \left< \frac{n_k!}{(n_k-l)!} \frac{n_j!}{(n_j-m)!}\cdots \right>
  = \left\{\left[\frac{x_k}{k}\right]^l \left[\frac{x_j}{j}\right]^m
    \cdots \right\} \frac{A!}{(A - \{lk + mj + \cdots\})!}
    \frac{Q_{A - \{lk + mj + \cdots\} } (\vec x)}{Q_A(\vec x)} . \label{nknj}
\end{eqnarray}
Using these correlations and fluctuations, we are able
to study intermittency and the fractal dimension in a nuclear
fragmentation process without any sampling problem.
In a Monte-Carlo type study such as in a percolation model,
the fluctuations can be enhanced due to the small number of samples.
The intermittency and fractal dimension appear to depend on
the parameter value $x$ indicating their temperature dependence.
Dividing the cluster size space $k = 1, 2, \cdots, A$ into $N = [A/L]$
bins of length $L$, where $[A/L]$ is the smallest integer with
$[A/L] \ge A/L$, the power moments
\begin{eqnarray}
 P_q(L) = \sum_{J=1}^{[A/N]} \left< \left( \sum_{j\in J} p(j) \right)^q
                    \right>    \label{pql}
\end{eqnarray}
gives a generalized Renyi entropy
\begin{eqnarray}
 S_q(L) = - \frac{\ln P_q(L)}{q-1}    \label{sql}
\end{eqnarray}
and a generalized dimension by setting the meassure
\begin{eqnarray}
 H_q(L) = P_q(L) \left(\frac{L}{A}\right)^{-(q-1) D_q} = 1 .  \label{hql}
\end{eqnarray}
The generalized (or anomalous) dimension is related to these power
moments $P_q(L)$ and the associated entropy by
\begin{eqnarray}
 D_q(L) = \left[\frac{1}{q-1}\right] \frac{\ln P_q(L)}{\ln [L/A]}
        = - \frac{S_q(L)}{\ln [L/A]} .
\end{eqnarray}
In the limit $q \to 1$, $D_1(L)$ becomes the information dimension
associated with the Shannon information defined by
\begin{eqnarray}
 S_1(L) = - \sum_{J=1}^{[A/N]} \left< \left[\sum_{j \in J} p(j)\right]
              \ln \left[\sum_{j \in J} p(j)\right] \right> .
\end{eqnarray}
The $L \to 0$ limit of $D_q(L)$ becomes the fractal dimension of the
distribution \cite{fractal}.
The size $L$ independence of $D_q(L)$ indicates the existence
of the self-similar fractal substructure in the distribution.
On the other hand, the order $q$ dependence of $D_q(L)$ originates
from a multifractal structure \cite{fractal}.
Analysis of nuclear emulsion data for Au at 0.99 GeV/A of Ref.\cite{auemuls}
shows the bin size $L$ independence
of the intermittency exponent $\alpha_q(L) = (q-1) D_q(L)$
and the generalized dimension $D_q(L)$ \cite{intm}.
This analysis also exhibits a $q$ independent behavior of $\alpha_q(L)$
and the $q$ dependence of $D_q(L)$.
Here we have considered the cluster distributions by the charge size
instead of the mass size.

\section{Emulsion data of various nuclei}

In order to compare the analysis of power moments among nuclei with different
sizes (the total charge number $Z$), we first define a dimensionless
variable $y$ such that
\begin{eqnarray}
 y = z/Z .
\end{eqnarray}
We then define the charge multiplicity density and the charge distribution
probability density as
\begin{eqnarray}
 n(y) &=& n_z Z ,  \\
 p(y) &=& p_z Z ~=~ m_z ~=~ n(y)y .
\end{eqnarray}
The $n(y)$ and $p(y)$ satisfy the normalization conditions of
\begin{eqnarray}
 \int_0^1 dy ~y n(y) = 1
   &=& \left[\sum_{z=1}^Z \frac{1}{Z}\right] \left[\frac{z}{Z}\right]
        \left[n_z Z\right] ,  \\
 \int_0^1 dy ~p(y) = 1
   &=& \left[\sum_{z=1}^Z \frac{1}{Z}\right] \left[p_z Z\right] .
\end{eqnarray}
The normalized cummulative charge can be defined as
\begin{eqnarray}
 M(y) = \int_0^y dy ~p(y)
   ~=~ \frac{M(z)}{Z} = \left[\sum_{i=1}^z \frac{1}{Z}\right] m_i .
\end{eqnarray}
where $M(z) = \sum_{i=1}^z m_i$ and $z = y Z$.
The charge distribution $p(y)$ and the staircase distribution
(the cummulative charge distribution) $M(y)$ averaged over the
canonical ensemble are compared for various nuclei in Fig.1 and Fig.2.
Considered here are the nuclear emulsion data \cite{auemuls,emulsdt} of
$^{84}_{36}$Kr at 1.52 GeV/A, $^{131}_{\ 54}$Xe at 1.22 GeV/A,
$^{197}_{\ 79}$Au at 0.99 GeV/A, and $^{238}_{\ 92}$U at 0.96 GeV/A.

These figures show that the mean behavior of the charge
distribution has a fast fall in the small elements region and a large flat
region at intermediate sizes. A rise at large elements also appears
for Kr and Au. According to the simple fragmentation model of Section 2
\cite{intm,otherx,general,frgft},
the fast drop at small clusters is the result of the fragmentation
of a hot system (large $x$), and the rise at large clusters
occurs when the system breaks up at low temperature (small $x$).
When $x = 1$, the charge distribution $p(y)$ is uniform.
To fit these data using the $x$-model of Section 2,
mixing \cite{intm} of at least two $x$ values is required,
one small value ($x \lsim 1$) and another larger value ($x >> 1$).
The fragmentation at high temperature may occur through a primary
fragmentation process, and the delayed subsequent fragmentation
processes would be a low temperature process.
A large participant created in a central collision would be a hot system
and a large spectator in a peripheral collision becomes a cold system.
A large system with low temperature may also be created by
a cooling process of the hot participant through evaporation.

To look at the mixed characteristics of the emulsion data with
various $x$ values further, we define a total
multiplicity distribution density $P(\mu) = P(M)Z$, where $\mu = M/Z$.
The $P(M)$ is the probability of partitions having
the total multiplicity $M$ in the canonical ensemble. Then
\begin{eqnarray}
 \int_0^1 d\mu P(\mu) ~=~ 1 ~=~ \sum_{M=1}^Z P(M)
   ~=~ \left[\sum_{M=1}^Z \frac{1}{Z}\right] \left[P(M) Z\right] .
\end{eqnarray}
We can also define a dimensionless spacing density as
\begin{eqnarray}
 P(s) = P(m_z) Z ,  \hspace{1cm} s = m_z/Z  .
\end{eqnarray}
This probability density satisfies the normalization condition
\begin{eqnarray}
 \int_0^1 ds P(s) ~=~ 1 ~=~ \sum_{m_z=1}^Z P(m_z)
   ~=~ \left[\sum_{m_z=1}^Z \frac{1}{Z}\right] \left[P(m_z) Z\right] .
\end{eqnarray}
Here $P(m_z)$ is the probability for a spacing value $m_z = z n_z$
in the canonical ensemble.
The multiplicity distribution $P(\mu)$ and the spacing distribution $P(s)$
are shown in Fig.3 and Fig.4 for various nuclei.

Fig.3 shows the multiplicity distribution.
The large scatter over the whole
range of possible multiplicity values, $0 \le \mu \le 1$,
indicates that the emulsion data may be formed through a combination of
fragmentation processes at many different temperatures.
According to the simple $x$-model, the total multiplicity at a given
temperature or with a fixed $x$ value exhibits a peaked distribution
with the position and the width of the peak depending on the value
of $x$ and $Z$ \cite{general}.
For small clusters, the multiplicity distribution at fixed $x$ follows
a Poisson distribution
\begin{eqnarray}
 P(n_k) = \frac{{\bar n_k}^{n_k}}{n_k!} e^{-\bar n_k}
\end{eqnarray}
quite well \cite{intm}.

The spacing distributions of Fig.4 resemble more a Wigner distribution
(dash-dotted line)
\begin{eqnarray}
 P_W(s) = \frac{2}{\pi\sigma} \left(\frac{s}{\sigma}\right)
                        e^{-(s/\sigma)^2/\pi}      \label{wspac}
\end{eqnarray}
rather than a Poissonian distribution (dashed line)
\begin{eqnarray}
 P_P(s) = \frac{1}{a} e^{- s Z/a} = \frac{1}{a} e^{- m_z / a} . \label{pspac}
\end{eqnarray}
In quantum chaos,
the Wigner distribution in the level spacing indicates a chaotic behavior.
This behavior might also indicate a mixed processes in the emulsion data.
However, for the spacing distribution of these data, the form (solid line) of
\begin{eqnarray}
 P_G(s) = \frac{s Z}{b^2} e^{-s Z/b} = \frac{m_z}{b^2} e^{-m_z/b}
                       \label{gspac}
\end{eqnarray}
fits better than the Wiger distribution.
The Poisson distribution of Eq.(\ref{pspac}) and the form of Eq.(\ref{gspac})
have a longer tail than the Wigner distribution of Eq.(\ref{wspac}).
Long tail in the spacing distribution indicates the existence of fragmentation
events with the charge distribution peaked at a single $z$ value.
Two extreme cases contributing to the long tail are the vaporization of
a nucleus into $z = 1$ elements
and a peripheral collision having large spectators.
A large cluster produced in a fragmentation at a low temperature
would also contribute to the long tail.

Changing the bin size $L$ into a dimensionless bin size parameter $r = L/Z$,
the power moments $P_q(L)$ and the meassure $H_q(L)$ remain unchanged.
We can see this as follows:
\begin{eqnarray}
 P_q(L) &=& \sum_{J=1}^{[A/L]} \left< \left(\sum_{j \in J}
                p(j)\right)^q \right>                    \nonumber   \\
        &=& \sum_{J=1}^{[A/L]} \left< \left( \left[\sum_{j \in J}
                \frac{1}{Z} \right] \left[p(j) Z \right] \right)^q \right>
                           \nonumber   \\
        &=& \sum_{J=1}^{[A/L]} \left< \left( \int_{y \in J} dy ~p(y)
                              \right)^q \right>        \nonumber   \\
        &=& P_q(L/Z) = P_q(r) ,    \label{pqr}  \\
 H_q(L) &=& P_q(L) \left(\frac{L}{Z}\right)^{-(q-1)D_q(L)}   \nonumber   \\
        &=& \left[\sum_{J=1}^{[A/L]} \left(\frac{L}{Z}\right)^{D_q}\right]
   \left< \left(\frac{\sum_{j \in J} p(j)}{(L/Z)^{D_q}} \right)^q \right>
                    \nonumber   \\
        &=& \left[\sum_{J=1}^{[A/L]} \left(\frac{L}{Z}\right)^{D_q} \right]
            \left< \left(\frac{\left[\sum_{j \in J} \frac{1}{Z}\right]
               \left[p(j) Z\right]}{(L/Z)^{D_q}} \right)^q \right>
                     \nonumber   \\
        &=& \left[\sum_{J=1}^{[A/L]} r^{D_q} \right]
            \left< \left(\frac{\int_{y \in J} dy ~p(y)}
                   {r^{D_q}} \right)^q \right>
                     \nonumber   \\
        &=& P_q(r) r^{-(q-1)D_q(r)} = H_q(r) .   \label{hqr}
\end{eqnarray}
Thus the power moments $P_q(r=L/Z) = P_q(L)$ and the generalized
dimension $D_q(r) = D_q(L)$ can be compared directly as shown in
Figs.5 -- 10.

Comparing Figs.5 -- 8 with Figs.1 and 2, we can explicitly see the
effects of both the fluctuations and correlations which are included
in $P_q(r)$ but not in the mean distribution of $p(y)$ and $M(y)$.
The dashed curves, which were the best fits for the mean behavior
of Fig.1 and Fig2, no longer give a good description of
the power moments of Figs.5 -- 8.
For Au and U data (Fig.7 and Fig.8), a fit with three different $x$ values
(thick solid line) fits better than one with two $x$'s (dash-dotted line)
for $q = 3$ even though they were somewhat similar for $q = 2$ case.
Figs.9 and 10 exhibit a scale invariant behavior for large nuclei
(Au and U) and a size dependence for small nuclei (Kr and Xe).
Fig.10 shows a linear behavior of the power moments $P_q(r)$ for Xe and U
with less linearity for the Kr and Au data.
The later nuclei have a rise at large elements in the charge distribution
$p(y)$ of Fig.1.

Scale invariance can be characterized by a power law behavior $r^\alpha$
of the power moments or an $r$ independence in the generalized
dimension $D_q(r)$. This size $r$ independence of the intermittency
exponent $\alpha(r) = (q-1) D_q(r)$ or in the generalized dimension $D_q(r)$
is related to a scale invariance of the distribution.
Shown in Fig.11 is the generalized dimension
calculated by
\begin{eqnarray}
 D_q(r) = \left[\frac{1}{q-1}\right] \frac{\ln P_q(r)}{\ln r}
        = \left[\frac{1}{q-1}\right] \frac{\ln P_q(L/Z)}{\ln(L/Z)} .
                \label{dqr}
\end{eqnarray}
We see that $D_q(r)$ is linear in $r = L/Z$ for small bins sizes.
For light nuclei (Kr and Xe), $D_q(r)$ increases linearly until
$r \approx 1/2$ while, for heavy system (Au and U), $D_q(r)$ exhibits
a saturating behavior after $r \approx 1/4$.
For bin sizes larger than $r \approx 1/2$, $D_q(r)$ decreases due to
the finite size of $Z$, i.e., the finite size effect on binning the system.
The linear fits for the dimension $D_q(r)$ using a form
\begin{eqnarray}
 D_q(r) = D_q(L) = C_q Z r + D_q(0) = C_q L + D_q(0)   \label{dqrlin}
\end{eqnarray}
with $L = r Z$ are shown in Fig.11 with thin lines.
The parameter values of $C_q$ and $D_q(0)$ are given in Table 1.
For small nuclei, the linear fit is quite
good until the bin size becomes $r \approx 0.5$ except for $r = 1/Z$.
The large increase in the dimension at $r = 1/Z$ from its value
at $r = 2/Z$ may have an explanation in the even-odd oscillating
behavior in Fig.1 of $p(y)$.
We also list in Table 1 the dimension calculated by
\begin{eqnarray}
 D_q'(r=L/Z) = \left[\frac{1}{q-1}\right]
         \frac{\log [P_q(L)/P_q(L+1)]}{\log [L/(L+1)]}   \label{dqrp}
\end{eqnarray}
as in Ref.\cite{intm}.
This dimension is much larger than the dimension calculated by
Eq.(\ref{dqr}) as expected from Fig.9.
This large result may also be related to the even-odd oscillating
behavior of $p(y)$.

The dimensions $D_q(r)$ for $r = 1/Z$ and $2/Z$ given in Table 1
or shown by Fig.11 exhibit a dependence on the order $q$.
This $q$ dependence is related
with a multifractal structure in fractal theory \cite{fractal}.
However, this dependence might also be related to the mixed charateristics
(two or more $x$'s) of the multifragmentation processes.
In the simple $x$ model,
the dimension $D_q(L)$ is $q$ independent for large $x$ values
where the power moments have a power law behavior \cite{intm}.
Except for Au, which has an approximately $q$ independent intermittency
exponent $\alpha_q = (q-1) D_q'(1) \approx 0.2$,
all nuclei considered here exhibits $q$ independence in
$\alpha_q/q$ rather than the generalized dimensions or the
intermittency exponents.
The values of $\alpha_q/q = (q-1) D_q'(1)/q$ are about
0.14 for Kr, 0.22 for Xe, and 0.16 for U.
Power moments of the mean charge distribution, which are
defined by Eq.(\ref{pqr}) with $<p(y)>$ instead of $p(y)$,
shows a $q$ indenpendence of $\alpha_q/q = (q-1) D_q(r) / q$
for small $x < 1$ (see Table III of Ref.\cite{intm}).

\section{Conclusion}

We have analyzed nuclear multifragmentation phenomena using
power moments of the charge distribution for the nuclear emulsion
data of $^{84}_{36}$Kr at 1.52 GeV/A, $^{131}_{\ 54}$Xe at 1.22 GeV/A,
$^{197}_{\ 79}$Au at 0.99 GeV/A, and $^{238}_{\ 92}$U at 0.96 GeV/A.
The mean charge distribution and the multiplicity distribution
of these data show that the emulsion data at about 1 GeV/A energy
may be a multiprocess phenomena.
The multiprocesses that generates the emulsion data could be the break
up of a hot system and a delayed fragmentation of an excited system
with low temperature.
The even-odd oscillating behavior in the mean charge distribution could
indicate some effects from the mean field interation at this energy region.

The charge spacing distribution of these data for small spacings
follows the Wigner distribution
rather than the Poisson distribution indicating that the process may have
some underlying chaotic mechanism. This behavior may be related with
the multiprocess characteristics of the data.
However, the spacing distribution for large spacings
has an exponential long tail similar to a
Poisson distribution rather than a Gaussian tail of a Wigner distribution.

The bin size $r$ independence of $D_q(r)$ indicates a scale invariance
of the fragment distribution.
The data analyzed in this paper shows a scale invariance for large nuclei
such as Au and U, but a size dependence for small nuclei such as
Kr and Xe. Similar trends have been seen in our simple $x$ model.
The multiplicity distribution of clusters with a given size exhibits
Poisson distribution for a large system (large $Z$), and a non-Poissonian
distribution for a small system due to the sum rule constraint of having
a fixed $Z$ or $A$ \cite{intm}.
The power moment analysis of the charge distribution also exhibits
characteristics of a multistep process.
The $q$ dependence of the generalized dimension $D_q(r)$ might
originate from mixed temperatures in the fragmenting system.

An interesting problem would be to analyze a multifragmentation process
for a system having one equilibriated temperature. For this analysis,
we need to identify all the prompt fragments produced by the primary
break up of a hot uniform nucleus formed through a central collision.
The fragments in emulsion data
that we have analyzed seemed to be produced through multistep processes.
We expect that the prompt fragmentation of a uniform hot nuclear system
might exhibit a $q$ independence in $D_q(r)$.

\acknowledgments

We would like to thank Prof. P.L. Jain for letting us use his
unpublished data of Kr, Xe, and U.
One of us (SJL) would like to thank
the hospitality offered him at the Department of Physics and Astronomy,
Rutgers University.
This research was supported in part by the Kyung Hee University
under grant number A-92-04 and in part by an N.S.F. grant (grant number
NSF89-03457).

\begin{figure}
\caption{The probability density for the charge distribution in
the scaled charge variable $y = z/Z$.
The solid circles connected by thin solid lines are the data.
The dashed and dash-dotted curves are the fit with
two $x$'s and the solid line is the fit with three $x$'s as listed in
Table 1.}
\end{figure}

\begin{figure}
\caption{Staircase behavior of the cummulative charge distribution
with $y$. The curves are the same as in Fig.1.}
\end{figure}

\begin{figure}
\caption{The probability density associated with the total
multiplicity distribution in the scaled
multiplicity variable $\mu = M/Z$. The curves are the same as in
Fig.1.}
\end{figure}

\begin{figure}
\caption[ ]{The distribution for the level spacing density in
the scaled spacing $s = m_z/Z$.
The dash-dotted line is the Wigner distribution of Eq.(\ref{wspac}) and
the dashed line is the Poisson distribution of Eq.(\ref{pspac}).
Also shown by the solid line is the fit with
the form of Eq.(\ref{gspac}).
The parameter values are given in Table 1.}
\end{figure}

\begin{figure}
\caption[ ]{Power moments for $^{84}_{36}$Kr at 1.52 GeV/A.
The curves are the same as in Fig.1. The thin solid curve is the data.}
\end{figure}

\begin{figure}
\caption[ ]{Power moments for $^{131}_{\ 54}$Xe at 1.22 GeV/A.
The curves are the same as in Fig.5.}
\end{figure}

\begin{figure}
\caption[ ]{Power moments for $^{197}_{\ 54}$Au at 0.99 GeV/A.
The curves are the same as in Fig.5.}
\end{figure}

\begin{figure}
\caption[ ]{Power moments for $^{238}_{\ 92}$U at 0.96 GeV/A.
The curves are the same as in Fig.5.}
\end{figure}

\begin{figure}
\caption[ ]{Power moments for order $q =$ 2, 3, 4, \& 5 in a double
log scale.}
\end{figure}

\begin{figure}
\caption[ ]{Power moments using a linear scale.}
\end{figure}

\begin{figure}
\caption[ ]{Anomalous dimension. The solid line, the dashed line,
the dotted line, and the dash-dotted line are for $q = 2$, 3, 4, and 5
respectively. The thick lines are the data and the thin lines are the
linear fits with the parameter values given in Table 1.}
\end{figure}

\widetext
\begin{table}
\caption[ ]{Parameters used in various fits.
The parameters for the $x$ model fit is given in the form of
$c_{x_1} + c_{x_2} + c_{x_3}$ which means a mixture of $x$'s with
$c_{x_1}$ of $x = x_1$, $c_{x_2}$ of $x = x_2$, and $c_{x_3}$ of $x =x_3$.
The parameters $\sigma$, $a$, and $b$ for the multiplicity fit are
defined by Eqs.(\ref{wspac}), (\ref{pspac}), and (\ref{gspac}) respectively.
The linear fit of the generalized (anomalous) dimension is given in the
form of Eq.(\ref{dqrlin}). The dimension $D_q(L)$ of Eq.(\ref{dqr}) and
$D_q'(L)$ of Eq.(\ref{dqrp}) are listed for $L = 1$ in the first column
and $L = 2$ in the second column for each nucleus.}
\begin{tabular}{ccccc}
        &
    $^{84}_{36}{\rm Kr} (1.52 {\rm GeV/A})$ &
    $^{131}_{\ 54}{\rm Xe} (1.22 {\rm GeV/A})$ &
    $^{197}_{\ 79}{\rm Au} (0.99 {\rm GeV/A})$ &
    $^{238}_{\ 92}{\rm U} (0.96 {\rm GeV/A})$  \\
\tableline
 curve & & $x$ model fit \\
 -- -- & $0.50_{0.5} + 0.50_{80}$ & $0.39_{1} + 0.61_{100}$  &
        $0.66_{0.5} + 0.34_{50}$ & $0.60_{1.3} + 0.40_{92}$ \\
 -- $\cdot$ & $0.63_1 + 0.37_{100}$ & $0.58_2 + 0.42_{150}$  &
        $0.75_{0.5} + 0.25_{20}$ & $0.75_1 + 0.25_{30}$ \\
 --- & $0.62_{0.7} + 0.10_5 + 0.28_{100}$ & $0.47_1 + 0.15_5 + 0.38_{100}$  &
        $0.72_{0.5} + 0.15_5 + 0.13_{80}$ & $0.57_1 + 0.23_5 + 0.20_{100}$ \\
\tableline
    & & Multiplicity  \\
 $\sigma$  & 0.050 & 0.055 & 0.060 & 0.060 \\
 $a$ & 5 & 10 & 10 & 10 \\
 $b$ & 5 & 5 & 5 & 5 \\
\tableline
 $q$ & & \hspace{0.1cm}$D_q(r=L/Z)$ & \hspace{-2.5cm}$ = ~C_q L + D_q(0)$  \\
 2 & $0.0062L+0.120$ & $0.0028L+0.143$ & $0.0042L+0.124$ & $0.0031L+0.175$ \\
 3 & $0.0045L+0.099$ & $0.0024L+0.115$ & $0.0038L+0.098$ & $0.0027L+0.143$ \\
 4 & $0.0038L+0.083$ & $0.0018L+0.098$ & $0.0032L+0.078$ & $0.0020L+0.123$ \\
 5 & $0.0035L+0.071$ & $0.0015L+0.085$ & $0.0028L+0.067$ & $0.0016L+0.108$ \\
    & & \hspace{1.0cm}$D_q(r=L/Z)$  &  \hspace{-2.0cm}for $L=1$, 2 \\
 2 & 0.157 \ \ 0.128 & 0.190 \ \ 0.144 & 0.136 \ \ 0.130 & 0.202 \ \ 0.181 \\
 3 & 0.126 \ \ 0.104 & 0.158 \ \ 0.117 & 0.103 \ \ 0.102 & 0.165 \ \ 0.148 \\
 4 & 0.107 \ \ 0.0889& 0.135 \ \ 0.0987& 0.0840\ \ 0.0848& 0.141 \ \ 0.127 \\
 5 & 0.0944\ \ 0.0782& 0.119 \ \ 0.0857& 0.0714\ \ 0.0732& 0.124 \ \ 0.112 \\
    & & \hspace{1.0cm}$D_q'(r=L/Z)$ & \hspace{-2.0cm}for $L=1$, 2 \\
 2 & 0.278 \ \ 0.0604& 0.407 \ \ 0.1142& 0.170 \ \ 0.0905& 0.318 \ \ 0.184 \\
 3 & 0.219 \ \ 0.0483& 0.351 \ \ 0.0892& 0.112 \ \ 0.0661& 0.258 \ \ 0.152 \\
 4 & 0.184 \ \ 0.0429& 0.309 \ \ 0.0757& 0.0803\ \ 0.0537& 0.217 \ \ 0.135 \\
 5 & 0.162 \ \ 0.0400& 0.278 \ \ 0.0669& 0.0617\ \ 0.0462& 0.188 \ \ 0.125 \\
\end{tabular}
\end{document}